\documentclass[letter]{jpsj2} 

\title{Bifurcations and Chaos in the 
Six-Dimensional Turbulence Model of Gledzer}

\author{\textsc{Makoto Umeki}\thanks{E-mail address: umeki@phys.s.u-tokyo.ac.jp}}

\inst{Department of Physics, 
Graduate School of Science, 
University of Tokyo, \\
7-3-1 Hongo, 
Bunkyo-ku, Tokyo, 113-0033 Japan}

\abst{
The cascade-shell model of turbulence with six real variables 
originated by Gledzer is studied numerically using Mathematica 5.1.
Periodic, doubly-periodic and chaotic solutions and the routes to 
chaos via both frequency-locking 
and period-doubling are found by the Poincar\'e plot of 
the first mode $v_1$. 
The circle map on the torus is well approximated by 
the summation of several sinusoidal functions. 
The dependence of the rotation number on the viscosity 
parameter is in accordance with that of the 
sine-circle map.
The complicated bifurcation structure and 
the revival of a stable periodic solution 
at the smaller viscosity parameter 
in the present model indicates that the turbulent state 
may be very sensitive to the Reynolds number. 
}

\kword{bifurcation, chaos, turbulence, shell model, Mathematica}

\begin{document}
\maketitle

Many of turbulent phenomena in motions of fluids are 
considered to be understandable by investigating 
ordinary differential equations which are models 
faithful to the Navier-Stokes (NS) equations. 
Recent progress of computer hardware and software, 
however, may still not be enough to make clear the 
route and the characteristics of turbulence.
Another prompt way to smatter turbulence is 
to omit the theoretical derivation of nonlinear 
models, to give up the full numerical simulation 
and to fall back on simpler models analogous to 
the NS equation. 
In this context, the cascade-shell models by Gledzer 
(1973)\cite{gled} and its complex version (Ohkitani and Yamada (OY 
1989)\cite{OY}, so-called the GOY model; see also 
Frisch (1995)\cite{Frisch}, Kato and Yamada (2003)\cite{KY}, 
Biferale(2003)\cite{Biferale1} for further references) 
have been studied numerically from the viewpoint of turbulence statistics.
In the present investigation, the bifurcation approach is made to 
the Gledzer's cascade-shell model of turbulence with six real 
variables. 
Biferale {\it et al.} (1995)\cite{Biferale2} has found the 
limit-cycle and torus attractors in the GOY model 
with a fixed viscosity along with loss of the stability of 
the Kolmogorov 1941 fixed point when the 
parameter $\epsilon$ related to the {\it helicity} exceeds 
critical values about 0.386 and 0.396, respectively. 
In contrast with the bifurcation found by Biferale {\it et al.} (1995)
\cite{Biferale2} 
in GOY interpreted as the Ruelle-Takens scenario, a detailed study of 
the circle map in the Gledzer model in this paper indicates 
that the bifurcation of torus attractors is in accordance with 
that of the sine-circle map.
Another similar study of the five-mode truncation 
model of the NS equation is made by Franceschini and Tebaldi (FT 1979)
\cite{FT}, where the period-doubling and the symmetry-breaking 
bifurcations are found but the quasi-periodic route 
is out of their result. 

The Gledzer's original model of real variables 
$v_i(t)$ for $i=1,\cdots, n$ is given by:
\begin{eqnarray}
\nonumber
\frac{dv_i}{dt}=\dot{v}_i = & c_{1,i}v_{i+1}v_{i+2}+c_{2,i}v_{i-1}v_{i+1}
\\
& +c_{3,i}v_{i-1}v_{i-2}- \nu k_i^2 v_{i} + f_{i}, 
\label{eq1}
\end{eqnarray}
where $v_i$ ($f_i$) is the velocity (forcing) of the $i$-th mode
in the space of a discretized wavenumber $k_i$. 
$\nu$ denotes the kinematic viscosity, 
the forcing is assumed to be time independent,
and $v_0=v_{-1}=v_{n+1}=v_{n+2}=0$.
In order to conserve the energy 
$E=\sum_{i=1}^{n}v_i^2$ 
and the enstrophy
$\sum_{i=1}^{n} k_i^2v_i^2$ 
in the case of $\nu=f_i=0$ 
as Gledzer (1973) required for the model to be 
analogous to the two-dimensional turbulence,
the coefficients of the nonlinear terms 
$c_{j,i}$ for $j=1,2,3$ and $i=1,\cdots, n$ 
need to satisfy the following relations:
\begin{equation}
c_{2,i+1}=-\frac{k_{i+2}^2-k_{i}^2}{ k_{i+2}^2-k_{i+1}^2} c_{1,i}, 
\label{eq2}
\end{equation}
\begin{equation}
c_{3,i+2}=-\frac{k_{i+1}^2-k_{i}^2}{ k_{i+1}^2-k_{i+2}^2} c_{1,i}.
\label{eq3}
\end{equation}
If the wavenumber and the nonlinear coefficient 
$c_{1,i}$ are selected as $k_i=c_{1,i}=k_0 q^i$,
the above relations become
\begin{equation}
c_{2,i}=-\beta k_{i-1}, \qquad 
\label{eq4}
\end{equation}
\begin{equation}
c_{3,i}= (\beta-1) k_{i-2},
\label{eq5}
\end{equation}
where 
\begin{equation}
\beta= 1+q^{-2}.
\label{eq6}
\end{equation}
However, we switch the parameters chosen by OY, 
$\beta=1/2$ and $q=2$ in (\ref{eq4}), (\ref{eq5}), $k_i$ and $c_{1,i}$, 
which do not satisfy (\ref{eq6}), model ling the 3D turbulence 
in the sense that the conservation law 
holds only for energy, not for enstrophy. 

Assuming that the forces are applied only on the 
first and second modes 
$f_1=f_2=1$ and $f_i=0$ for $i \ge 3$, 
the equation for the total energy with the forcing 
and the viscosity becomes
\begin{equation}
\dot{E}/2 = f_1 v_1+ f_2 v_2 - \nu \sum_{i=1}^n k_i^2 v_i^2.
\label{eq7}
\end{equation}
If we consider the sufficiently large values of 
$v_i$ such that $|v_i| > |f_i| /(\nu k_i^2)$ for 
the forced modes $i=1,2$, the right hand side 
becomes negative and the solution is 
proved to be finite.
The present model deals with real variables 
and therefore it is possible to reduce 
the dimensions of the model to the half 
with the same extent of the wavenumber, 
compared with the GOY model. 

Letting $k_0=2^{-4}$, the model becomes 
\begin{eqnarray}
\nonumber
\dot{v}_i= & 2^{i-5}v_{i+1}v_{i+2}-2^{i-7}v_{i-1}v_{i+1} \\
& -2^{i-8}v_{i-1}v_{i-2}- \nu 4^{i-2} v_{i} + f_{i}.
\label{eq8}
\end{eqnarray}
The present model possesses a single parameter $\nu$, 
changed between the range about $5 \cdot 10^{-2}$-$5 \cdot 10^{-3}$. 
Three ways of giving the initial condition are considered 
in order to examine multiple stable states. 
Case (I) is the origin; $v_i(0)=0$. 
Case (II) [or (III)] is the final computed values of the 
slightly smaller [or larger] value of the viscosity $\nu$. 

The Linux version of Mathematica 5.1 is used for the 
main part of the numerical integration. 
According to the manual of Mathematica, the general-purpose 
ODE solver developed by Hindmarsh (1983)\cite{Hindmarsch} 
is adopted in the {\it NDSolve} command. 
The numerical solution is constructed checking its local 
convergence up to the machine precision of order $10^{-16}$. 
The weakest point is that it consumes large memory. 
The number of the modes is fixed as $n=6$ in the main part 
of this study. The computer has 1GB memory, its CPU is 
AMD Athlon XP 2000+ and its operating system is Vine Linux 3.2.
A Fortran program with the fourth-order Runge-Kutta scheme 
with a fixed time step is also coded to check the 
bifurcation diagram in Figure 2 and no inconsistency 
is found between results by Mathematica and Fortran. 

The number $N(n)$ of fixed points (FPs) 
of (\ref{eq8}) including complex numbers 
are computed by Mathematica 
and shown as $N(4)=5$, $N(5)=9$, $N(6)=25$, $N(7)=29$, 
$N(8)=61$, $N(9)=129$ and $N(10)=177$, 
but the number of real fixed points is only 
5 for $n=4,5$ and 3 for $6 \le n \le 10$ at $\nu=10^{-2}$.

Figure 1 shows the bifurcation structure of the FPs; 
the stable (unstable) FPs are denoted by thick (thin) curves.
The Mathematica program is coded such that the 
number of modes is arbitrary. 
The fact that the bifurcation structure is 
quantitatively the same between $n=6$ and 8 suggests 
the sufficiency of the six-dimensional system 
in the region $0.005<\nu <0.05$. 
The number of stable FPs is 1 for 
$\nu>0.03983 (3)$, $0.03482> \nu >0.02059 (5)$, 
$0.01287> \nu >0.01202 (5)$, 
2 for $0.02058 > \nu > 0.01288 (5)$ and 
0 for $0.03982> \nu > 0.03483 (3)$, $0.01201>\nu$. 
The number of total FPs is shown in the above parenthesis; 
3 for $0.05>\nu > 0.03483$, $0.01167> \nu >0.00967$, 
5 for $0.03482> \nu >0.01168$, $0.00968 > \nu >0.00791$, 
7 for $ 0.00790> \nu >0.00355$ and 
9 for $0.00354> \nu$. 
There is a tendency that as $\nu$ decreases, 
the number of real FPs increases. 

Figure 2(a) shows the bifurcation diagram of attractors 
by plotting values of $v_1$ at the local maxima of $v_{1}$ 
($\dot{v}_{1}=0$ and $\ddot{v}_1 < 0$)
between $t=9 \cdot 10^3$ and $10^4$, 
actually computed by the interpolation of 
three adjacent points with the time interval $dt=0.01$.
The initial condition is Case (I). 
The curve in the center of Fig.2(a) is therefore 
due to the stable FP. 

The parameter region including the stable doubly-periodic 
state is examined in detail using the initial condition 
of Case (II) and (III).  
The stable doubly-periodic solutions are confirmed to be 
generated through the supercritical Hopf bifurcation 
at about $\nu=0.03834$ as shown in the enlarged 
Figure 2(b) and Figure 2(c). 
The doubly-periodic solutions coexist with the 
periodic solutions with period $6T$, on which the 
incomplete period-doubling bifurcation occurs 
as seen in Fig. 2(b). 
Here, the period $mT$ means that the periodic solutions 
give $m$ points on the present Poincar\'e plot.
The torus breakdowns at about $\nu=0.03785$ 
where the $6T$-periodic solution is still stable. 
Fig. 2(c) shows that there are windows of 
frequency-locking, which is one of the 
remarkable features of the sine-circle map. 
According to the current resolution in Figure 2(d), 
the $6T$-periodic solution changes chaotic suddenly 
at about $\nu=0.03724$, indicating the intermittency route. 

The left side of the chaotic parameter region 
in Fig. 2(a) is enlarged in Figure 2(e). 
Periodic windows are also observed among the 
chaotic solutions. 
A sequence of period-doubling bifurcations 
(Feigenbaum 1978)\cite{Feigenbaum} of orbits
is also observed in Figure 2(f), although it is 
in fact a revival of the periodic solution from 
chaos since the horizontal axis denotes the viscosity 
and the inviscid limit $\nu \rightarrow 0$ turns to the left 
in the Fig. 2(f). 
Numerically computed examples of attractors projected 
on the $(v_1,v_2,v_3)$ space for various values of 
$\nu$ are shown in Figure 3. 

In order to examine the circle map of the quasi-periodic solution,  
the pair $(\hat{v}_1^{(n)}, \hat{v}_1^{(n+1)})$ is considered, 
where $\hat{v}_1^{(n)}= v_1^{(n)}-<v_1^{(n)}>$,
$v_1^{(n)}$ is the $n$-th plotted value of $v_1$ and $<\cdot >$ 
denotes the time average.
The points lie on a closed curve in the Poincar\'e section, 
showing the evidence of the doubly-periodic torus motion 
in the original 6D space. 
Denoting the 2$\pi$-normalized argument of 
$\hat{v}_1^{(n)}+i \hat{v}_1^{(n+1)}$ by 
$\theta_n$ $(0\le\theta_n <1)$, 
the one-dimensional map 
$\theta_{n+1}=f(\theta_n)$ can be constructed. 

In many studies of the circle map, the sinusoidal 
function is considered, but strictly speaking, 
the function is likely to deviate from it in real 
chaos systems of ODEs.
We seek an approximation of the circle map 
of the form 
\begin{eqnarray}
\nonumber 
& \theta_{n+1}=
f(\theta_{n})= \theta_n+\Omega \\
 & +\displaystyle
\sum_{j=1}^{M} \frac{K_j}{2\pi j}
\sin [2\pi j( \theta_{n}-\theta_0^{(j)})], \ \ {\rm mod \ \ 1}
\label{eq9}
\end{eqnarray}
by applying the {\it FindFit} command of Mathematica 
to the numerically obtained data of the Poincar\'e 
plot with $M=10$. 
Figures 4(a-c) show the dependence of 
$K_j (j=1,\cdots , M)$ and $\Omega$ on the viscosity parameter. 
We observe the nonvanishing values of $K_2,K_4,K_6$
even at the onset of the stable torus; 
i.e. the second Hopf bifurcation. 

The numerical computation of the rotation number $\rho$ 
on the torus is also possible by counting 
the number $N_d$ of the decrease of $\theta_{n}$ due to the 
modulus 1 in Eq. (\ref{eq9}). A naive numerical approximation 
is $\rho=N_d/N$, where $N$ is the number of the total 
Poicar\'e plot, denoted by small dots in Figure 5.
A more elaborate method uses the power spectrum of $v_1(t)$. 
Choosing the suitable pair of two frequencies 
giving the maximum of the spectrum, we can identify 
the rotation number, denoted by larger dots in Fig.5. 
These two results are in agreement within the resolution 
of the numerical computation. 
The qualitative similarity is observed with that of 
the sine-circle map, showing the monotonic decrease 
of $\rho$ and the flat region, i.e. the evidence of the 
{\it incomplete} devil's staircase\cite{JBB}. 
Since our system has only one parameter, 
we just need to imagine the situation 
that two parameters $(\Omega, K)$ in the sine-circle 
map are changing simultaneously, as shown in Figures 4. 

In summary, the two different routes to chaos 
coexist in the single Gledzer's model of six 
real variables. As noted by E. Ott\cite{Ott} on page 199, 
the quasi-periodic route to chaos does not necessarily 
imply the bifurcation from double periodicity 
directly to triple periodicity or chaos, 
so-called the Ruelle-Takens scenario. 
The study of the sine-circle map indicates 
that frequency-locking leads to periodic solutions, 
which may become chaos through the period-doubling route 
or the other ones like symmetry-breaking or intermittency. 
The author has tested the Langford\cite{Langford} 
equation, well known to have torus attractors, 
and finds a similar scenario suggested by Ott\cite{Ott}.

According to Figures 4 and 12 in OY\cite{OY}, 
10 positive Lyapunov exponents have been found 
for the 48-dimensional model of 24 complex variables 
and the dimension of the corresponding attractor 
should be very large. 
Their parameters $\nu=10^{-8}$ and $f=(1+i)\times 5\times 10^{-3}$ 
correspond to the normalized viscosity $\nu'=\nu/{\rm Re} f
= 2\times 10^{-6}$ in our model with $f=1$. 
Their computation time $t=400$ leads to the normalized time 
$t'=({\rm Re} f) t = 2$ in our case with $n=6$. 
The computation of Lyapunov exponents of our model, 
as well as motions of eigenvalues of the Jacobian, 
have been made but is not shown in this Letter. 
The triply-periodic motion has not yet identified in our model. 
It is still open whether the low-dimensional chaos 
shown in this study is connected or not with the state of 
turbulence which is considered to be independent of 
the parameter $\nu$, 
as we increase the number of $n$ and decrease $\nu$. 
From the viewpoint of the bifurcation, it requires 
careful analysis but the gap will be filled as 
the power of the computer increases. 

The author is grateful to Prof. T. Yamagata for encouragement, 
to Prof. H. Sakai for permission to use the file server, 
to Prof. W. F. Langford for sending me the related 
reprints on the Langford equation, and to Prof. L. Biferale 
for letting the author know his work before proofreading. 

\

Figure 1. Bifurcation diagram of the fixed points (FPs) 
between $5\cdot 10^{-2}>\nu >5\cdot 10^{-3}$.
Stable (unstable) FPs are shown by thick (thin) curves.

\

Figure 2. Bifurcation diagram of attractors shown by values 
$v_1$ at the local maxima of $v_1$ ($\dot{v}_1=0$ and $\ddot{v}_1<0$). 

(a) The range of the parameter $\nu$ around 
$5\cdot 10^{-2}-5\cdot 10^{-3}$ and the initial condition 
is the origin.

(b) The parameter is $0.0372<\nu<0.0409$ and the initial condition 
is Case(II) and (III) such that the multiple stale states of the 
double periodicity and the 6$T$-periodicity are clearly captured. 
20000 points are selected randomly for minute drawing. 

(c) The enlarged Figure of (b) at $0.0378<\nu<0.03834$ showing the 
second supercritical Hopf bifurcation, the doubly-periodic and 
frequency-locked periodic solutions (especially, 
11:9, 17:14, 23:19 and 29:24 resonance), in addition to another 
stable 6$T$-periodic solution. 

(d) The parameter is $0.0372<\nu<0.0384$. The chaotic solution 
emerges suddenly at about $\nu=0.03724$, indicating 
the intermittency route. 

(e) Chaotic attractors with periodic windows at 
$0.006<\nu<0.012$.

(f) The enlarged figure of (e) at $0.006734<\nu<0.00674$, 
showing the evidence of the period-doubling bifurcation. 

\

Figure 3. Projections of attractors on the ($v_1$, $v_2$, $v_3$) space 
at $\nu=$ (a) 0.0395, (b) 0.0379, (c) 0.0378, (d) 0.037, (e) 0.0355, (f) 0.0118, 
(g) 0.0103, (h) 0.00883, (i) 0.0076, (j) 0.0074, (k) 0.0064 and (l) 0.006. 
(a) and (l) are the periodic solution of the period $T$, (k) $2T$, 
(f) $3T$, (h) $4T$ and (g), (c) $6T$. (b) is the doubly periodic 
and (d), (j) are the chaotic solutions, respectively.
(e) appears to be close to the heteroclinic orbit.

\

Figure 4. The dependence of the parameters in the circle map 
$K_i$ with even $i$ (a), odd $i$ (b) and $\Omega$ (c)
on the viscosity $\nu$.

\

Figure 5. The numerically obtained dependence of 
the rotation number $\rho$ on the viscosity $\nu$.
The small (large) dots denote $\rho$ by  
the method counting the number of 
decrease of $\theta_n$ due to the modulus 
(computing the ratio of two frequencies 
giving the maximum of the power spectra of $v_1$). 


\ \ 
\begin{center}
\scalebox{0.9}{\includegraphics{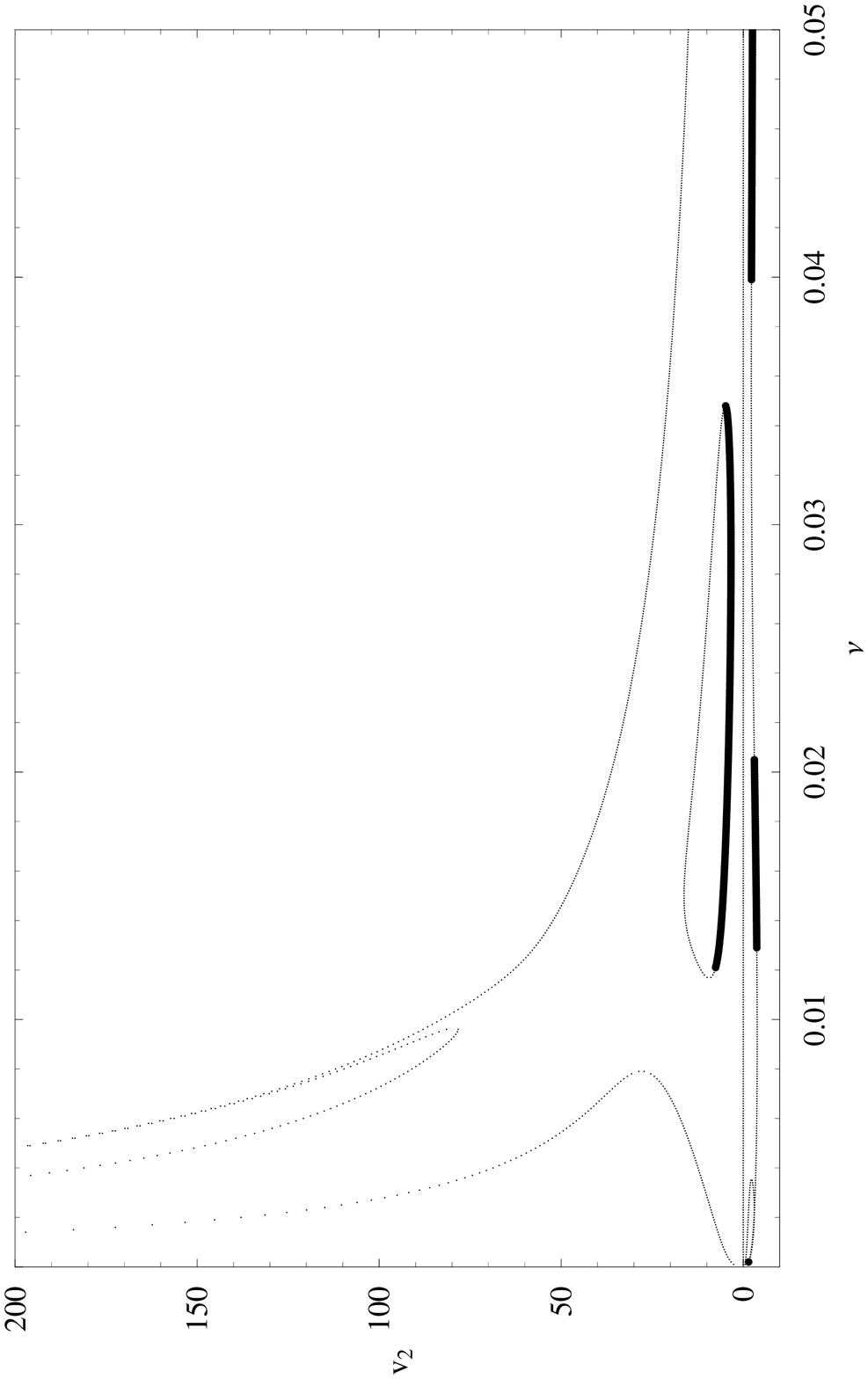}}
Figure 1
\end{center}

\begin{center}
\scalebox{0.8}{\includegraphics{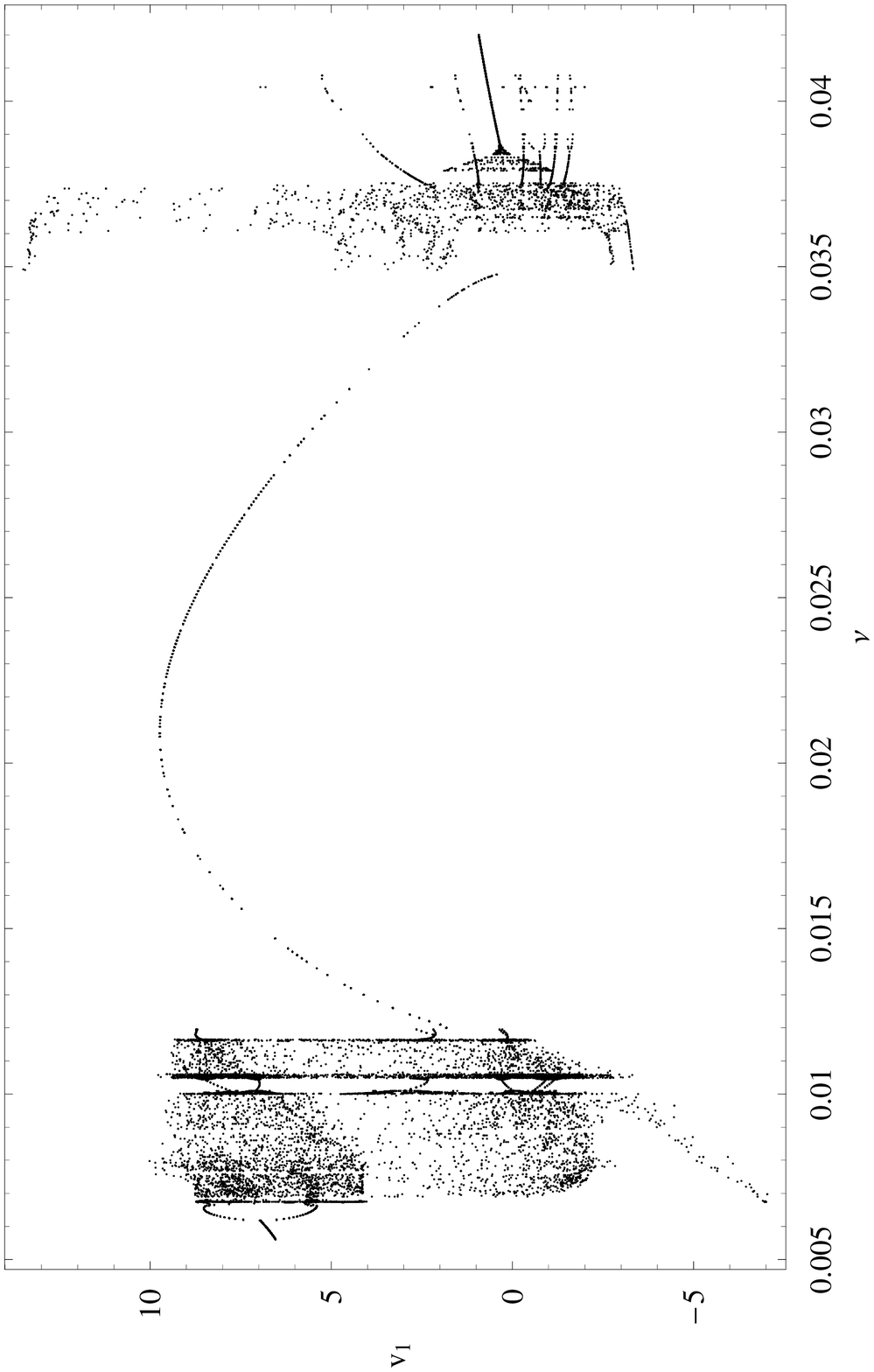}}

Figure 2a
\end{center}

\begin{center}
\scalebox{0.8}{\includegraphics{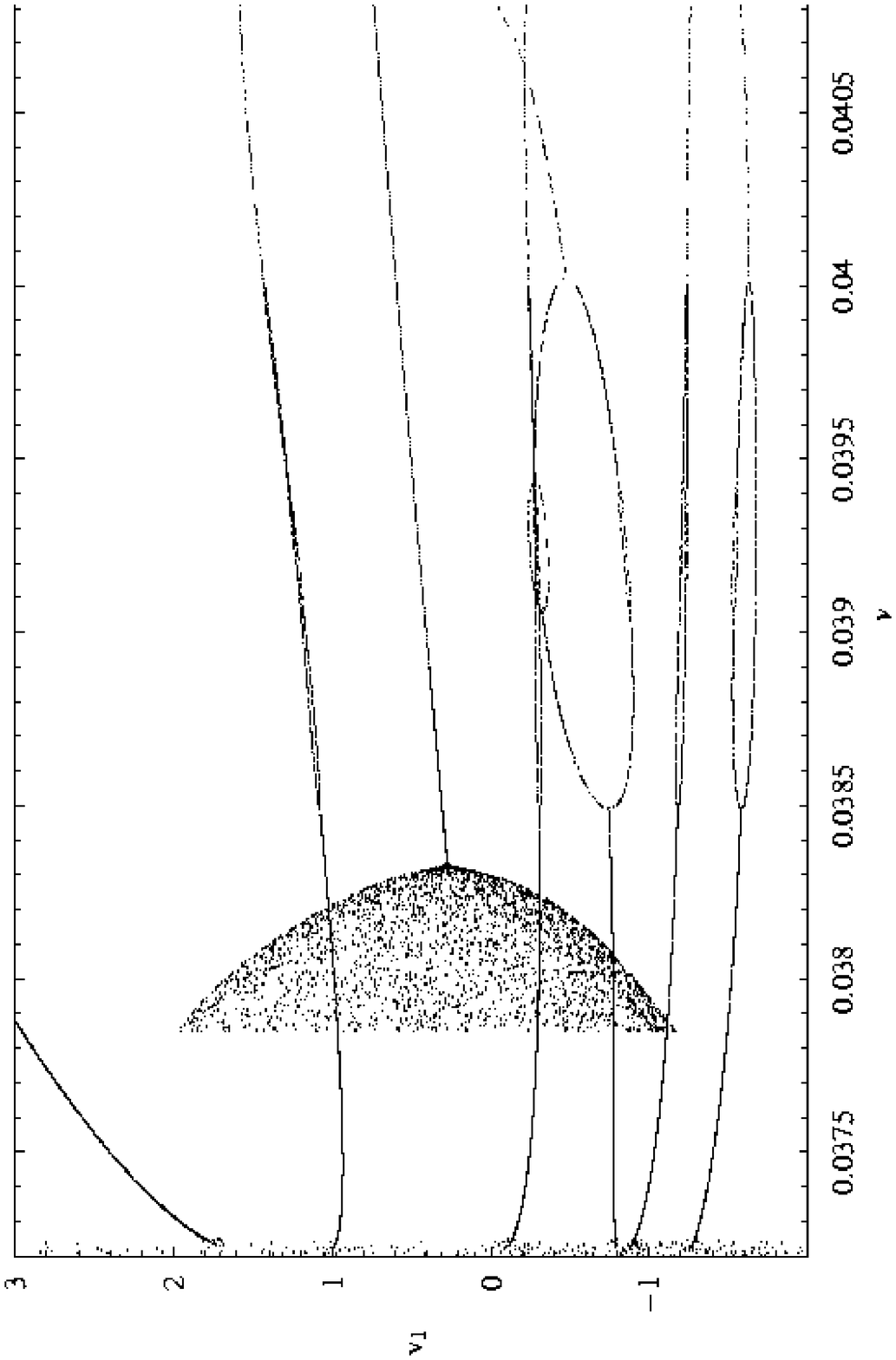}}
Figure 2b
\end{center}

\begin{center}
\scalebox{0.8}{\includegraphics{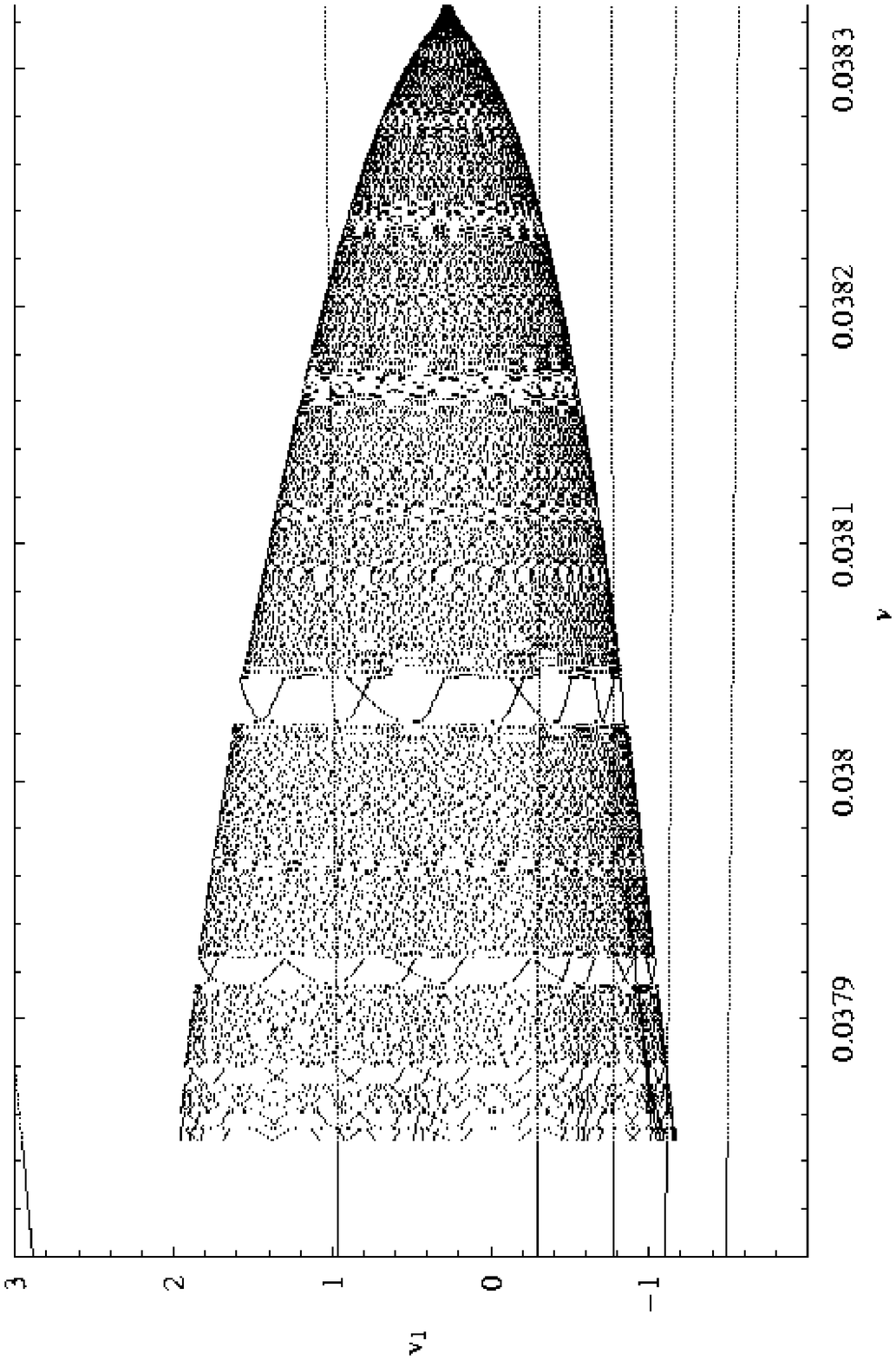}}
Figure 2c
\end{center}

\begin{center}
\scalebox{0.8}{\includegraphics{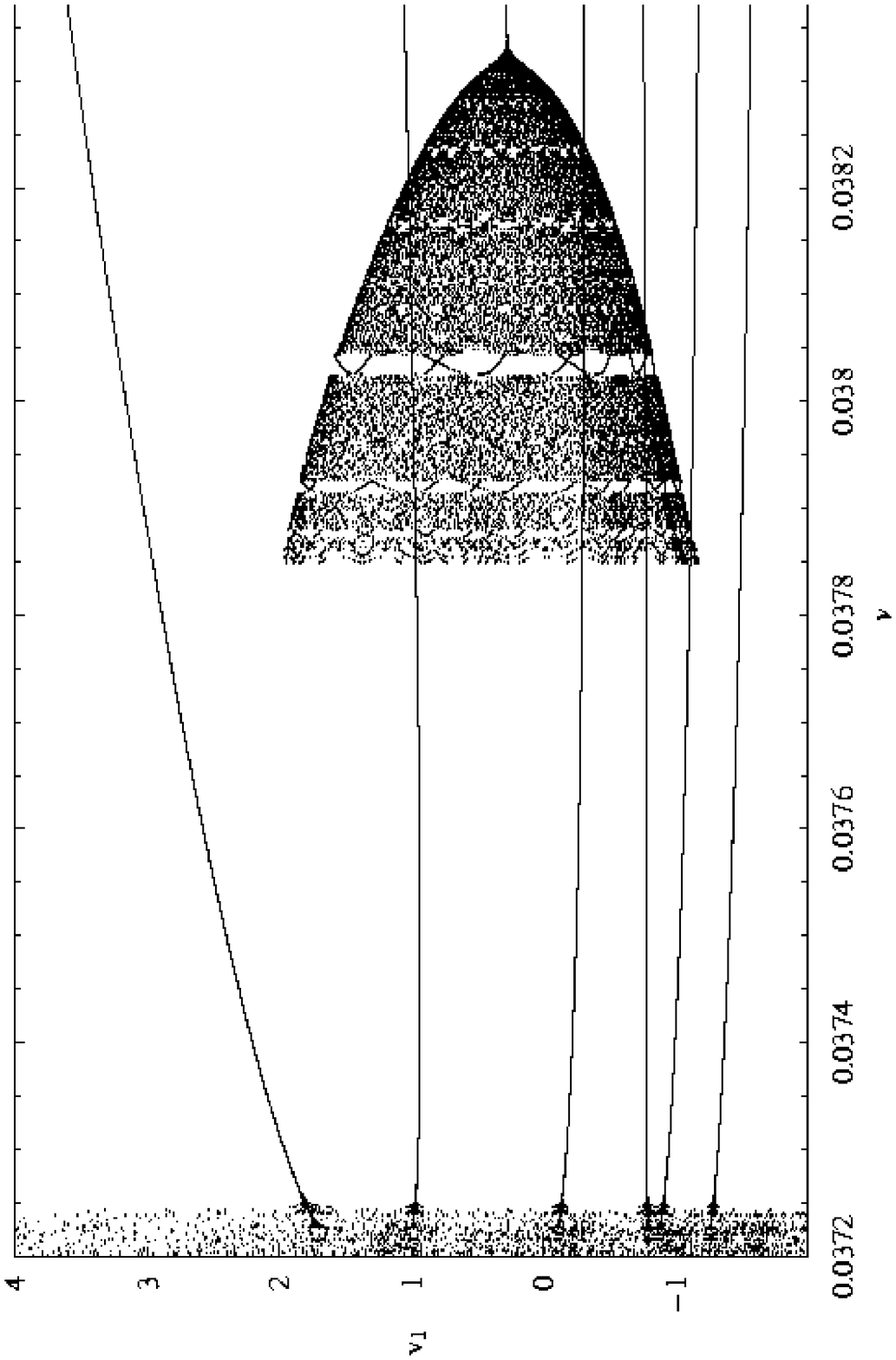}}
Figure 2d
\end{center}

\begin{center}
\scalebox{0.8}{\includegraphics{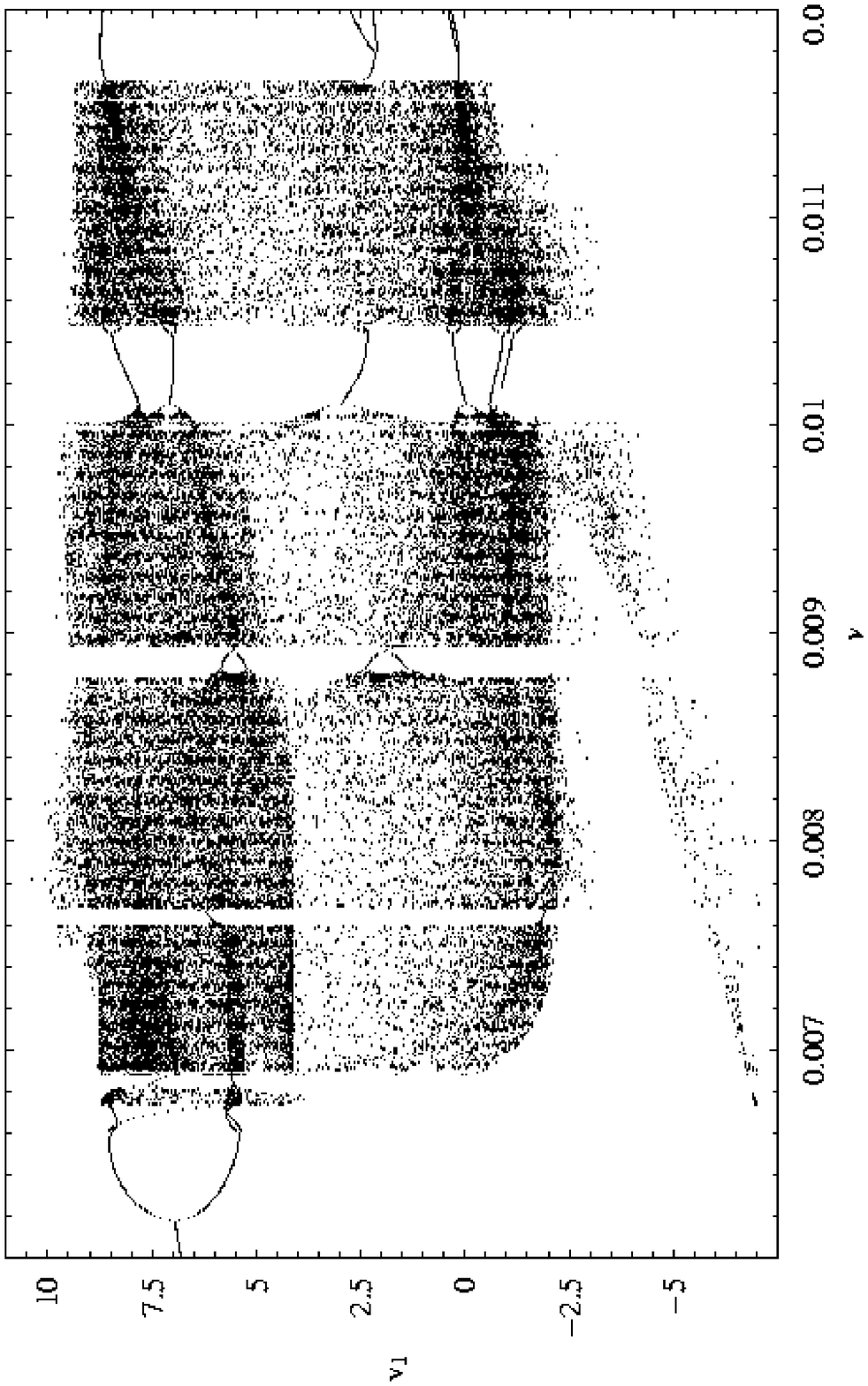}}
Figure 2e
\end{center}

\begin{center}
\scalebox{0.8}{\includegraphics{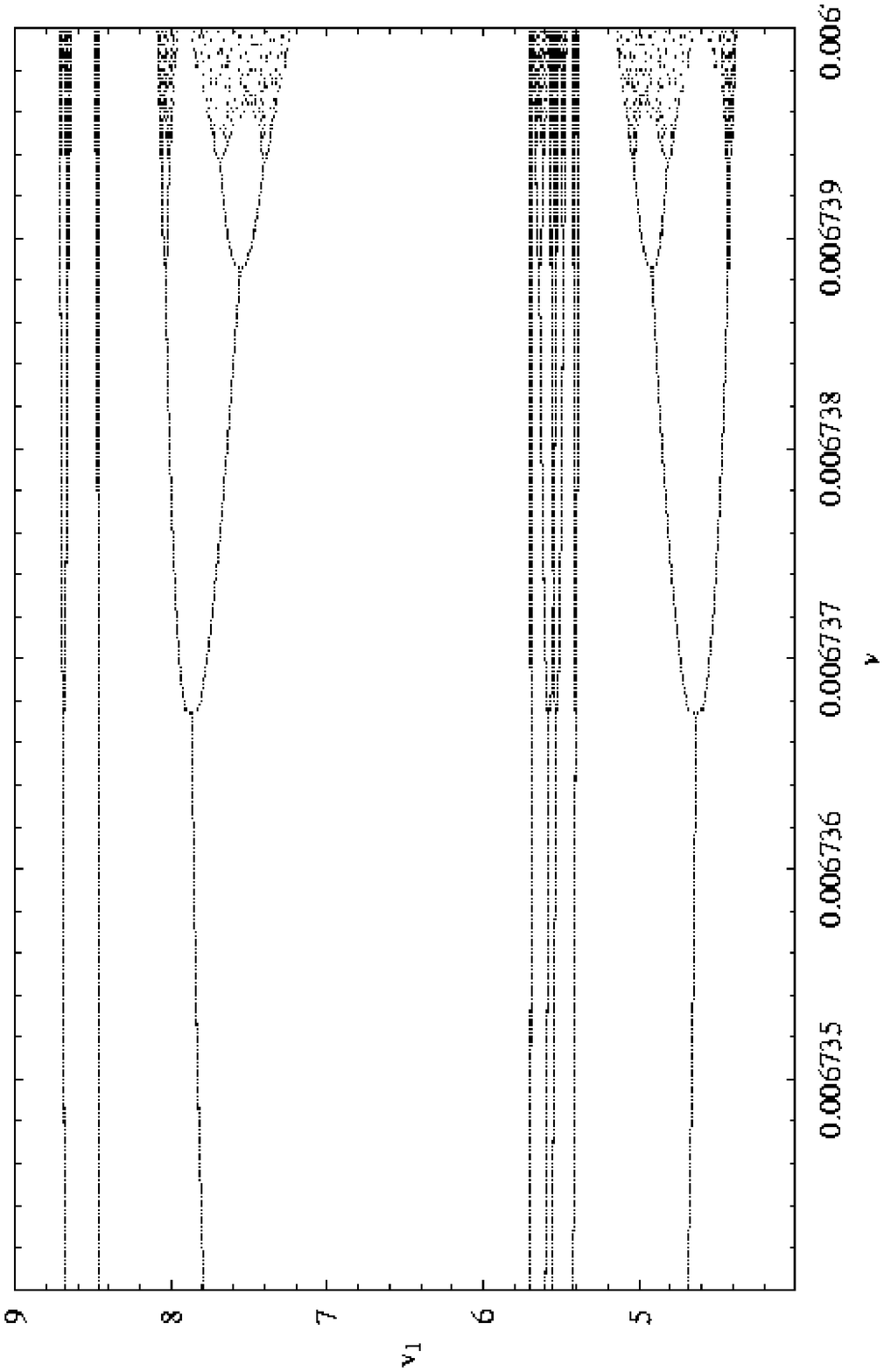}}
Figure 2f
\end{center}

\begin{center}
\scalebox{0.8}{\includegraphics{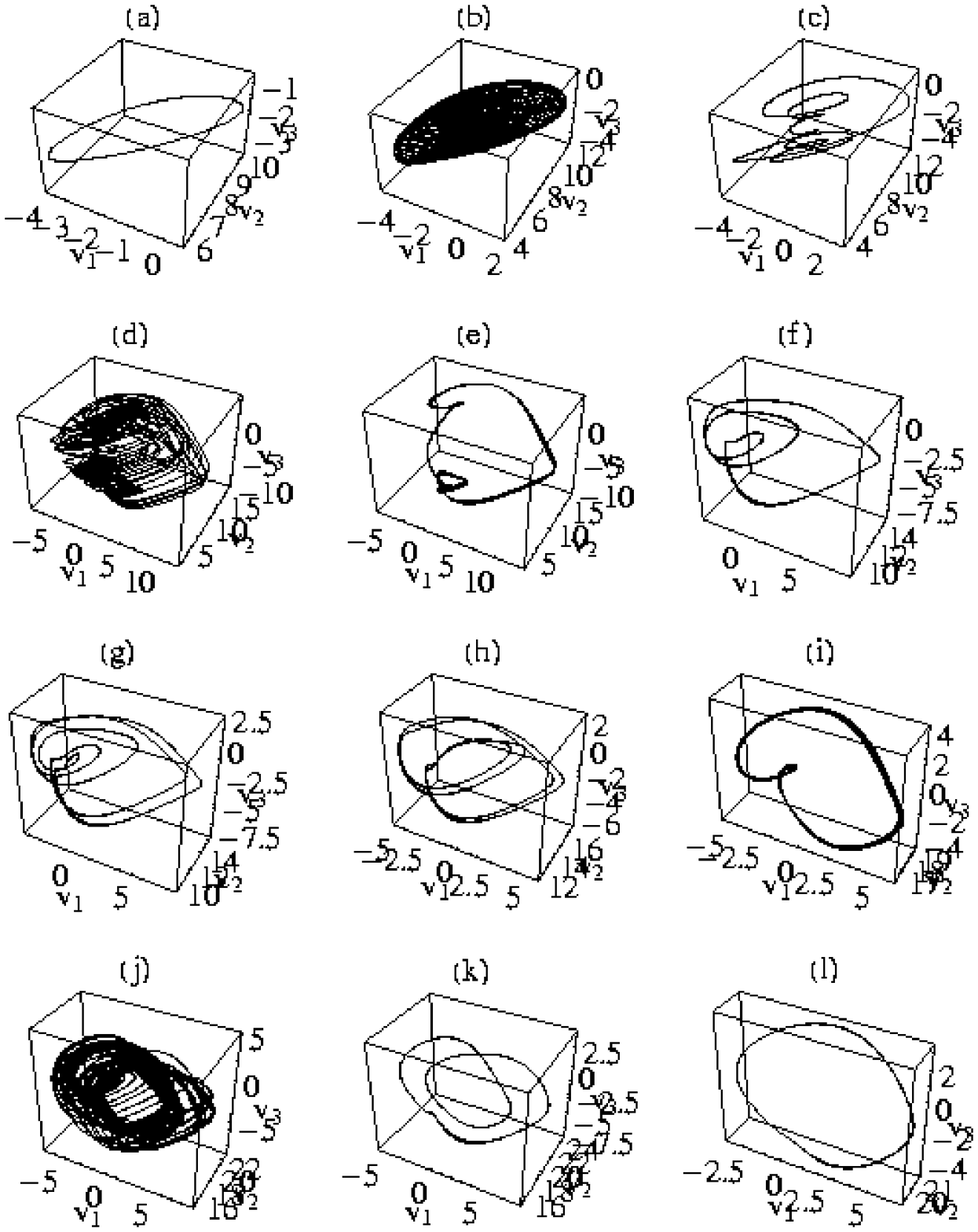}}
Figure 3
\end{center}

\begin{center}
\scalebox{0.88}{\includegraphics{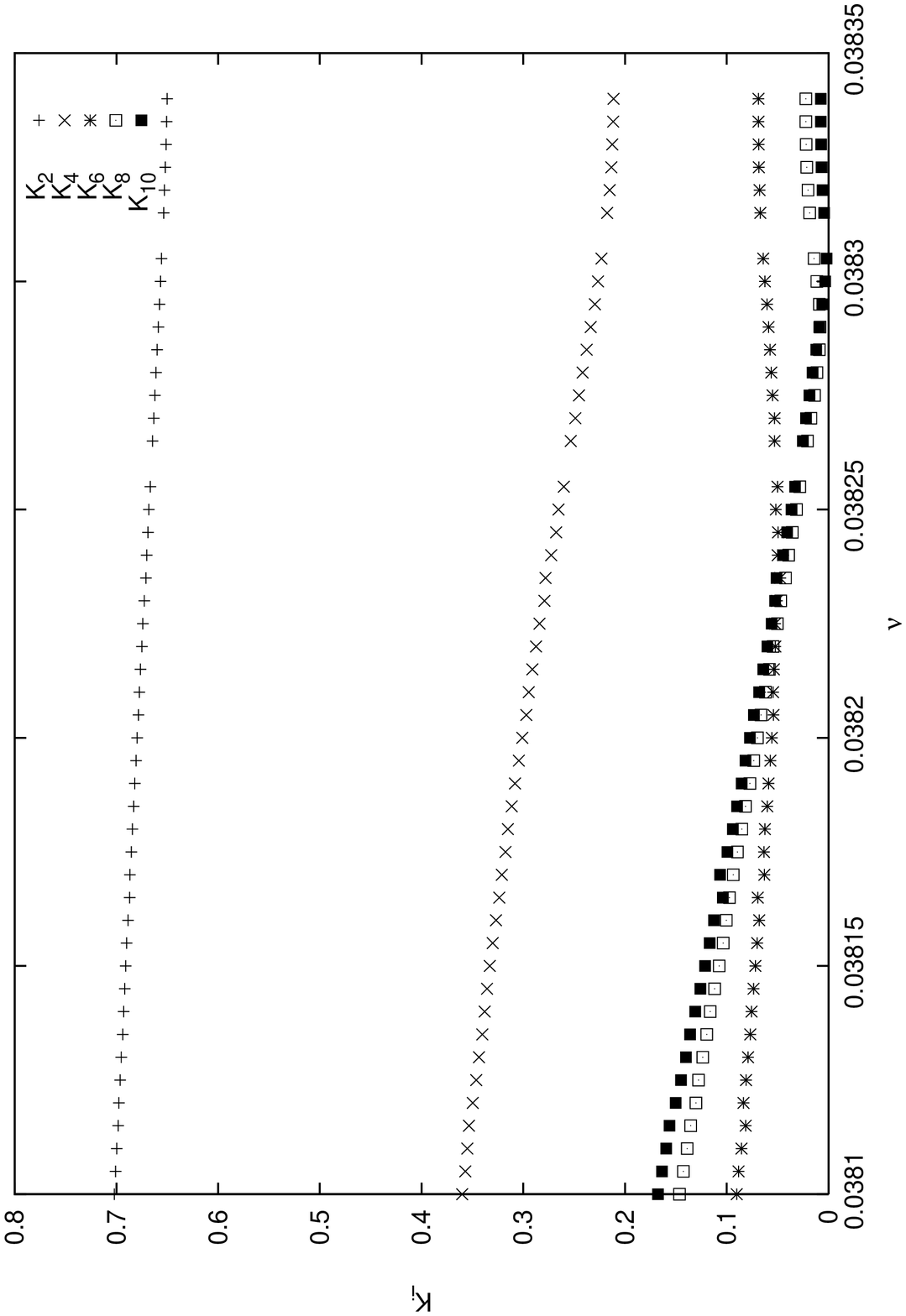}}
Figure 4a
\end{center}

\begin{center}
\scalebox{0.88}{\includegraphics{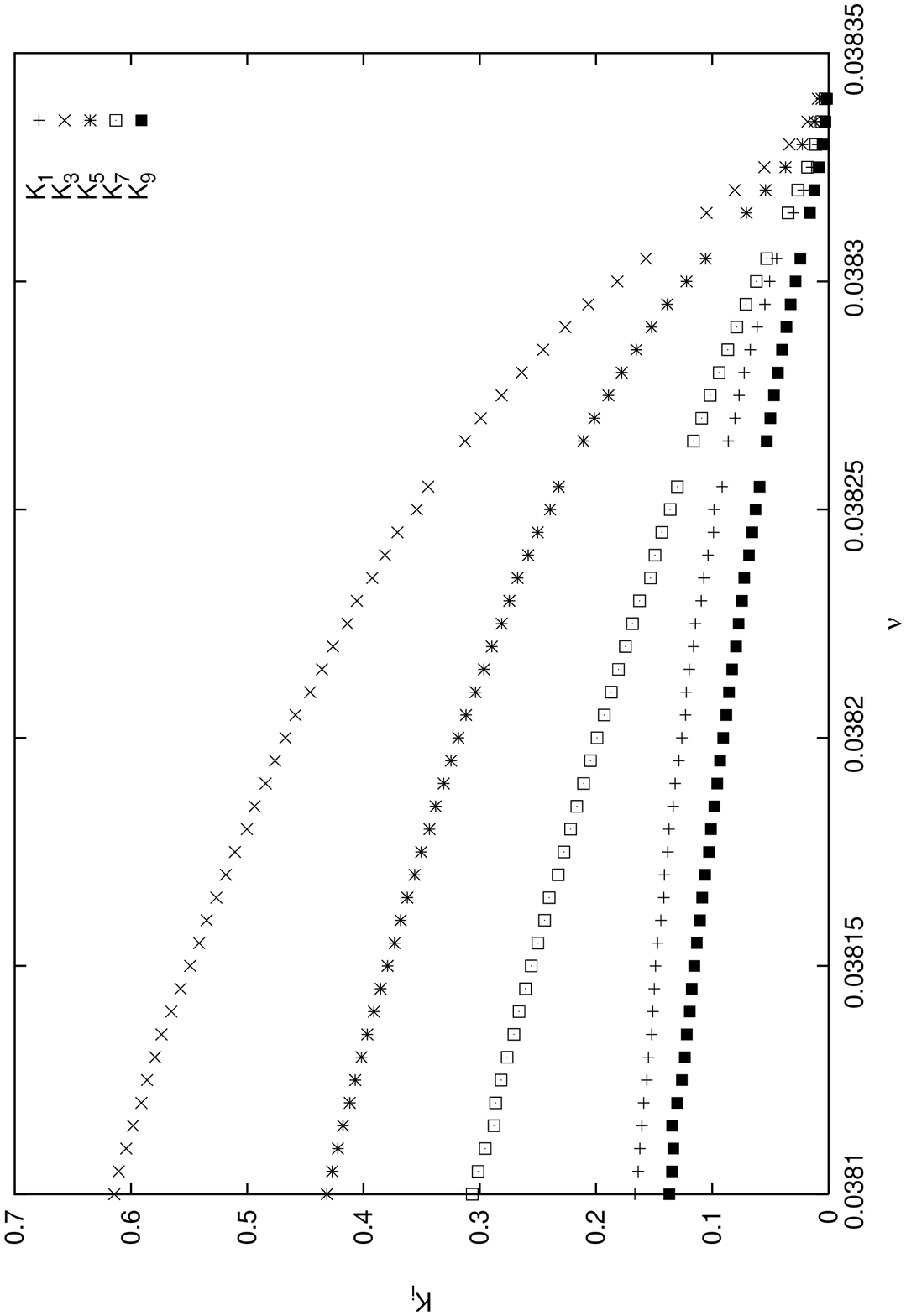}}
Figure 4b
\end{center}

\begin{center}
\scalebox{0.88}{\includegraphics{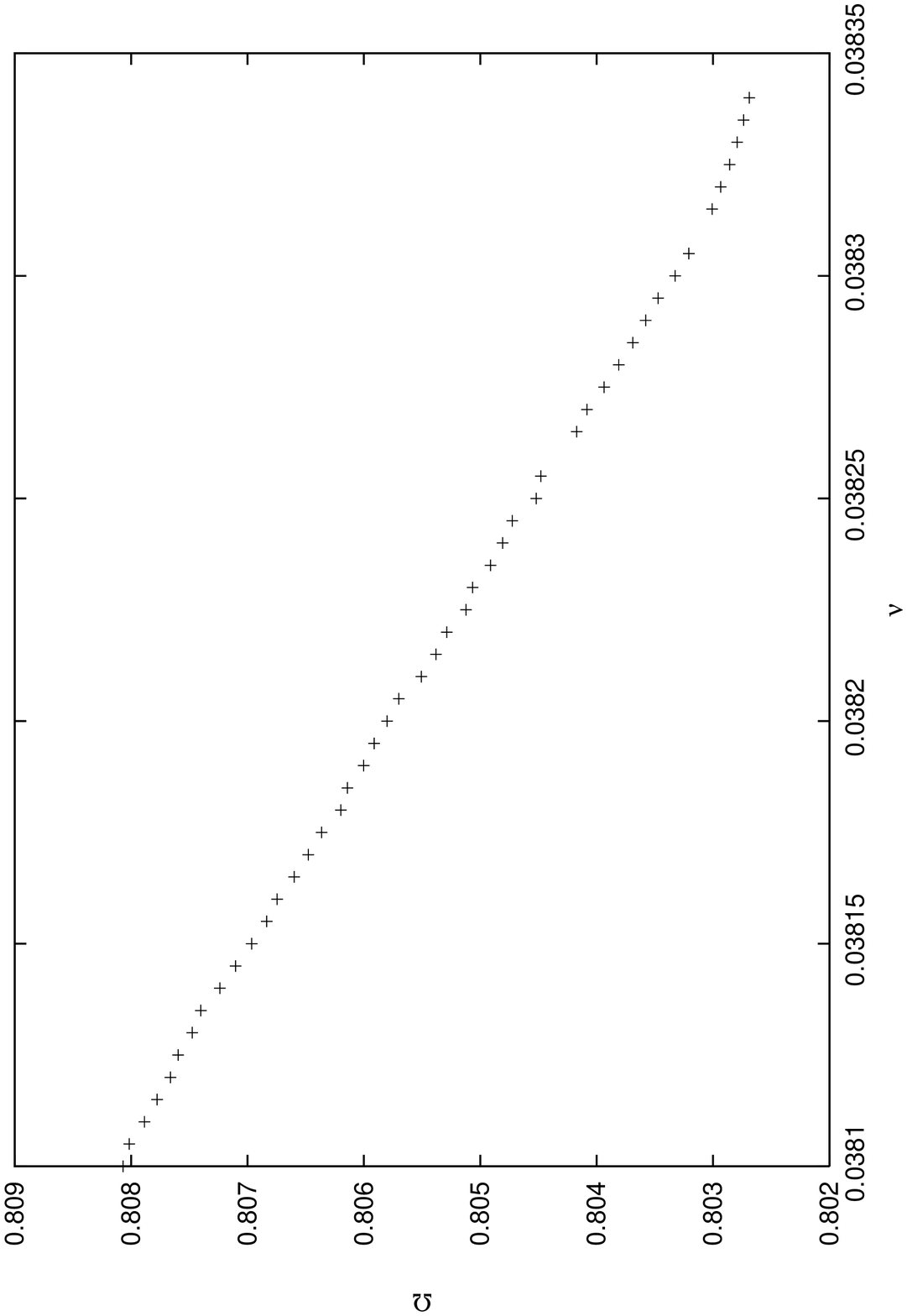}}
Figure 4c
\end{center}

\begin{center}
\scalebox{0.88}{\includegraphics{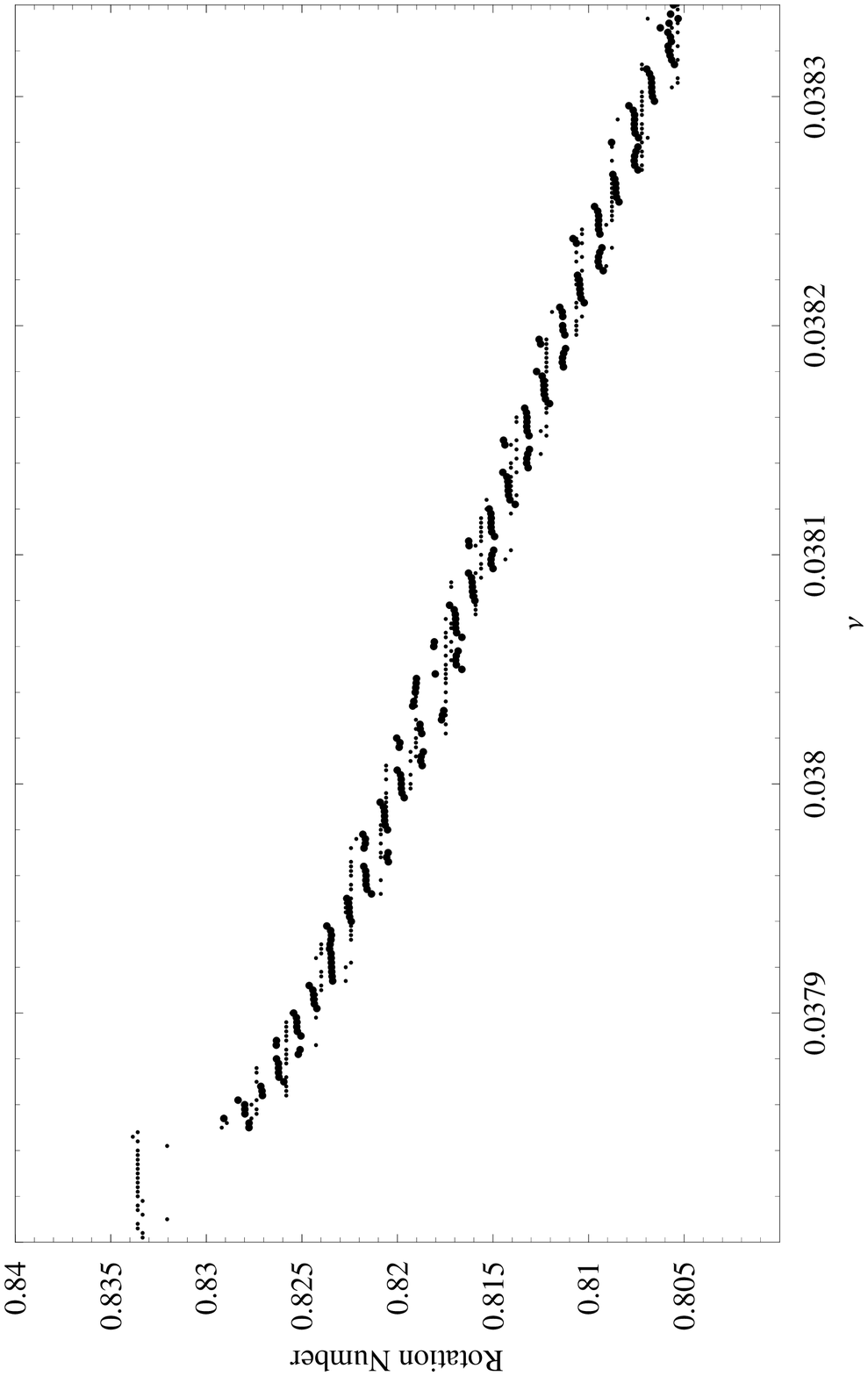}}
Figure 5
\end{center}

\end{document}